\documentstyle[preprint,aps,epsf]{revtex}
\newcommand{\be}{\begin{equation}}
\newcommand{\ee}{\end{equation}}
\newcommand{\bea}{\begin{eqnarray}}
\newcommand{\eea}{\end{eqnarray}}
\newcommand{\bd}{\begin{displaymath}}
\newcommand{\ed}{\end{displaymath}}

\begin{document}
\title{NON-UNIVERSAL GAUGINO MASSES IN SUPERSYMMETRIC SO(10)}
\author{Nidal Chamoun}
\address{Institute of Theoretical Physics, Chinese Academy of Sciences,\\ 
P.O. Box 2735, Beijing 100080, China and\\
Department of Physics, Higher Institute for Applied Sciences and 
Technology,\\ P.O. Box 31983, Damascus, Syria}
\author{Chao-Shang Huang, Chun Liu and Xiao-Hong Wu}
\address{Institute of Theoretical Physics, Chinese Academy of Sciences,\\ 
P.O. Box 2735, Beijing 100080, China}
\maketitle
\begin{abstract}
  We consider SUSY SO(10) models in which SUSY breaking occurs via an 
F-term which does not transform as an SO(10) singlet.  This results in 
non-universal GUT-scale gaugino masses leading to a different pattern of 
sparticle masses from what is expected in the minimal supergravity model 
(mSUGRA).  We study three breaking chains of SO(10) down to the standard 
model through SU(4)$\times$SU(2)$\times$SU(2), SU(2)$\times$SO(7) and 
`flipped' SU(5) achieved by the representations ${\bf 54}$ and 
${\bf 210}$ which appear in the symmetric product of two SO(10) adjoints. 
We examine the phenomenological implications of the different boundary conditions 
corresponding to the different breaking chains and present the sparticle 
spectrum.
\end{abstract}
\hspace{1.1cm}Keywords: SO(10) grand unification, Supersymmetry, Gaugino 
mass 

\newpage

\section{Introduction}
\label{Introduction}

Grand unification theories (GUT) are among the most promising models 
for physics beyond the standard model (SM).  Supersymmetry (SUSY) is 
necessary to make the huge hierarchy between the GUT scale and the 
electroweak scale stable under radiative corrections.  There are 
some experimental evidences for the SUSY GUT.  One is the apparent 
unification of the measured gauge couplings within the minimal SUSY SM 
(MSSM) at scale $M_{GUT}\sim 2\times 10^{16}$ GeV \cite{a}.  Another is 
the small neutrino masses extracted from recent observation of neutrino 
oscillations \cite{sk}, which imply that the next scale of new physics is 
the GUT scale.  

The most simple GUT model is the SU(5).  The next simple one is the 
SO(10) \cite{so10} which will be studied in this paper.  Whenever there 
is no intermediate scale of new physics between the GUT and electroweak 
scales, the gauge coupling unification is guaranteed. SO(10) models have 
additional desirable features over SU(5) ones.  All the matter fermions 
in one generation fit into one spinor representation of SO(10).  The 
representation contains the right-handed neutrino and, thus, provides an 
interesting framework in view of the small neutrino masses \cite{seesaw}.  
The R-parity can be automatically conserved as a consequence of some 
gauge symmetry breaking \cite{rparity}. Higgs fields can be put into any irreducible
representation (irrep) we want since Adler-Bell-Jackiw anomalies are 
absent for all representations of SO(10). In addition, SO(10) has more attractive subgroups, 
such as SU(5)$\times$ U(1), $SU(4)\times SU(2)\times SU(2)$ etc., thus it has more interesting 
breaking patterns than SU(5) has. Practically, it seems that the SUSY SU(5) is not favored by 
experiments \cite{su5}.
  
Because of SUSY breaking, the MSSM has over hundred soft parameters, 
such as gaugino masses, which are determined by the SUSY breaking 
mechanism.  The minimal supergravity model (mSUGRA) provides an 
attractive and economical framework to fix the soft parameters in the 
MSSM.  In mSUGRA, SUSY is broken in a ``hidden sector'', then 
gravitational-strengh interactions automatically transmit SUSY breaking 
to the ``visible sector'' which contains all the SM fields and their 
superpartners. Furthermore, one assumes that the 
K$\ddot{a}$hler potential takes a certain canonical form; as a result, 
all scalar fields get the same contribution $m_0^2$ to their squared 
scalar masses and all trilinear parameters have the same value $A_0$. 
In addition, one assumes that the gauge kinetic function is a function of the gauge singlet
so that gaugino masses have a ``universal'' value $m_{1/2}$ at 
scale $M_{GUT}$. The resulting weak scale spectrum of superpartners and 
their couplings can then be derived in terms of the SM parameters in 
addition to four continuous plus one discrete parameters $m_0$, 
$m_{1/2}$, $A_0$, $\tan{\beta}$ and sign($\mu$) provided that the radiative
breaking mechanism of the electroweak symmetry is assumed.  

However, these universal boundary conditions adopted in the mSUGRA are 
simple assumptions about the nature of high-scale physics and may remove 
some interesting degrees of freedom. Indeed, there exist interesting 
classes of mechanism in which non-universal soft SUSY breaking terms can 
be derived.  For instance, string-inspired supergravity or models in extra dimensions
 can lead to non-universality for SUSY breaking parameters at the string unification scale or
compactification scale.
\cite{brignole94and95}. There exists interesting phenomenology in SUSY models with
non-universal gaugino masses~\cite{hl}.

Non-universal gaugino masses may arise in supergravity models in which a 
non-minimal gauge field kinetic term is induced by the SUSY breaking 
vacuum expectation value (vev) of a chiral superfield that is charged under the 
GUT group \cite{anderson96andamundson96}.  The effect of non-singlet SUSY 
preserving vev on gaugino masses was studied in Refs. \cite{ellis85} and 
\cite{anderson96andamundson96}.  The boundary conditions for the 
gaugino masses have been worked out for the case of SU(5) GUT 
\cite{anderson96andamundson96} and their phenomenological implications 
have been investigated \cite{anderson2000}.
To our knowledge, there have not been studies of the non-universal 
gaugino masses resulting from SUSY breaking vev of SO(10) non-singlet chiral
superfields and the purpose of this paper is to provide just such a study.

The paper is organized as follows.  
In section II we present the  group-theoretical results which determine the 
boundary conditions for the gaugino masses coming from a condensation of 
F-component of a chiral superfield of a SO(10) non-singlet which is in the symmetric 
Kronecker  product of two SO(10) 
adjoints and we restrict our study to the lower dimensional representations 
$\bf 54$ and $\bf 210$. Each of these irreducible representations (irreps) leads to a proper pattern of 
non-universal gaugino masses depending on which breaking chain it leads to.  
For the irrep $\bf 54$ it can lead to two breaking chains from 
SO(10) down to the SM. One chain is through the phenomenologically 
interesting subgroup SU(4) $\times$ SU(2) $\times$ SU(2) $\equiv G_{422}$ 
corresponding to Pati-Salam model \cite{ps} and the other chain is through 
the subgroup SU(2) $\times$ SO(7). We determined the gaugino masses 
corresponding to these two chains irrespective of the Higgs multiplet used 
to break the intermediate subgroup to the SM. In accordance with the 
successful MSSM prediction of the gauge coupling unification, we assume 
that the breaking of the intermediate stages takes place also around 
$M_{GUT}$. As to the $\bf 210$ representation, it can lead to many breaking 
chains \cite{sato96,he90} and we chose to study the chain through the `flipped' 
SU(5) $\times$ U(1) \cite{flip} followed by a breaking via a $\bf 10$ 
representation of SU(5) contained in the spinor rep of SO(10) down to the SM.  Using these 
boundary conditions, one can use the renormalization group equations (RGEs) to deduce the weak scale 
values of the sparticle spectrum which we present in section III and we 
compare them to mSUGRA case. In calculating the spectrum we take into account  all constraints from the the present negative
searches of sparticles ( superpartners of SM particles and extra Higgs bosons) at collider experiments and $b\rightarrow s\gamma$ as well as the recent
data of the E821 experiment on the muon anomalous magnetic dipole moment.  In section IV we end up with concluding 
remarks.

\section{SUSY SO(10) GUTs with non-universal gaugino masses}

We discuss the non-universality of soft SUSY-breaking gaugino masses in SUSY SO(10) GUT.
 
In this class of models, non-universal gaugino masses are generated by at least a non-singlet chiral 
superfield $\Phi$ that appears linearly in the gauge kinetic function and whose auxiliary $F$ component 
acquires a vev~\cite{anderson96andamundson96,anderson2000}
\begin{eqnarray}
{\cal L}\supset \int d^2\theta f_{ab}(\Phi)W^aW^b + h.c. &\supset&
\frac{<F_{\Phi}>_{ab}}{M}\lambda^a \lambda^b
\end{eqnarray}
where the gauge kinetic function is $f_{AB}=f_0(\Phi_s)\delta_{AB}+\sum_{n}f_n(\Phi_s)
\frac{\Phi_{AB}^n}{M}+\ldots $ with M being a parameter of the mass dimension, $\Phi_s$ and $\Phi^n$ are the singlet and non-singlet chiral superfields respectively,
 the $\lambda^{a,b}$ are the gaugino fields and the $F_{\Phi}$ is the auxiliary field component 
of $\Phi$.

In conventional models of supergravity breaking, the assumption that only the singlet field $F_{\Phi_s}$ 
gets a vev is made so that one obtains universal gauge masses. However, in principle, the chiral superfield 
$\Phi$ which communicates supersymmetry breaking to the gaugino fields can lie in any representation found 
in the symmetric product of two adjoints
\begin{eqnarray}
({\bf 45}\times{\bf 45})_{symmetric}&=&{\bf 1}+{\bf 54}+{\bf 210}+{\bf 770}
\end{eqnarray}

where only ${\bf 1}$ yields universal masses. Thus the gaugino masses $M_a$ where the index $a=3,2,1$ 
represents the SM SU(3)$\times$SU(2)$\times$U(1) generators as a whole are, in general, 
non universal at the $M_{GUT}$ scale. 

In principle, an arbitrary linear combination of the above representations is also allowed and 
here we make two basic assumptions. The first one is that the dominant component of gaugino 
masses comes from one of non-singlet F-components. The second one is that the SO(10) gauge 
symmetry is broken down at the scale $M_{GUT}$ to an intermediate group $H$ by a non zero vev 
for the scalar component of the non-singlet superfield. In its turn, $H$ is subsequently broken 
down to the SM at the same scale $M_{GUT}$. Only the non-zero vev of the component of $F_{\Phi}$ 
which is `neutral' with respect to $H$ yields gaugino masses since $H$ remains unbroken after 
SUSY breaking. Depending on which breaking chain one follows down to the SM, ratios of gaugino 
masses $M_a$'s at $M_{GUT}$ are determined by group theoretical factors. We restricted our study 
to the lower dimensional representations ${\bf 54}$ and ${\bf 210}$ and
we discussed several
possible breaking chains in the following subsections.

Before we present our detailed discussions on gaugino masses, a remark is in place. According to the above
recipe that gives gaugino masses at tree level, the SUSY-breaking vev of the non-singlet 
superfield is also responsible for the breaking of the gauge symmetry. Because of the SO(10)
breaking down to H, there are heavy gauge supermultiplets which correspond to the broken generators
and receive masses of order of $m_{GUT}$. However, the SUSY-breaking effects proportional to
the vev of non-singlet F-component split the heavy gauge supermultiplets so that they behave
as messengers which communcate SUSY-breaking to the H gauge supermultiplet ( as well as the
quark and lepton supermultiplets ) by loop effects. The soft terms ( gaugino and squark
masses etc.) generated by the gauge-mediated mechanism with gauge messengers have been calculated in
ref. \cite{gr}. Applying their results to our case, the loop-induced soft terms can be neglected compared to those generated
at tree level (i.e. those discussed in the paper) if $M\sim M_{GUT}$
since they are proportional to $\frac{\alpha(m_{GUT})}{4\pi}$ which is about $3\times 10^{-3}$. In general, the size of M is
 model- dependent  and between $m_{GUT}$ and $m_{Planck}$ . For example, $M\sim m_{GUT}$ in the M-theory  on  $S^1/Z_2$,
$M\sim m_{string}\simeq 4\times 10^{17}$ GeV in the wekly coupled heterotic
$E_8\times E_8$ string theory, and $M\sim m_{Planck}$ in general supergravity
models. In the paper we limit ourself to the case of $M\sim m_{GUT}$.

\subsection{The representation $\bf 54$}
Looking at the branching rule for the GUT group SO(10) \cite{slansky81}, we see that the representation 
${\bf 54}$ can break it into several subgroups (e.g. $H=G_{422}\equiv SU(4)\times SU(2)\times SU(2), 
H=SU(2)\times SO(7), H=SO(9)$). Noting that the choice $H=SO(9)$ would lead to universal gaugino masses, we 
choose to study the following breaking chains. 

\subsubsection{$SO(10)\mapsto H=G_{422}\mapsto SU(3)\times SU(2)\times U(1)$}
The group $G_{422}$ corresponds to the phenomenologically interesting Pati-Salam model $SU(4)_C \times
SU(2)_R \times SU(2)_L$ where the lepton number is the fourth colour. The branching rule of the SO(10) represention
{\bf 54} to $G_{422}$ is
\be
{\bf  54} = ({\bf 20, 1, 1}) + ({\bf 6, 2, 2})  + ({\bf 1, 3, 3}) + ({\bf 1, 1, 1}). 
\ee
So at the first step of the breaking chain, we assume that the traceless $\&$ symmetric $2^{nd}$-rank tensor 
$\bf 54$ representation scalar fields have the non-zero vev
\begin{eqnarray}
<{\bf 54}>&=& v~ Diag(2,2,2,2,2,2,-3,-3,-3,-3)
\end{eqnarray}
where the indices $1,\ldots 6$ correspond to $SO(6)\simeq SU(4)$ while those of $7,\ldots 0$ (henceforth 
the index $0$ means $10$) correspond to $SO(4)\simeq SU(2)\times SU(2)$. To break $G_{422}$ down to the SM, 
one may simply choose the {\bf 16} Higgs fields since the branching rule of the rep. {\bf 16} to SM is 
\be
{\bf 16} = ({\bf 3, 2})_{1/3} +({\bf 3, 1})_{2/3}+({\bf \bar{3}, 1})_{-4/3} +({\bf 1, 2})_{-1}+
({\bf 1, 1})_2+({\bf 1, 1})_0,
\ee
where the number on the lower right denotes the quantum number Y of $U(1)_Y$.
The decomposition of the gauge (super) miltiplet {\bf 45} of SO(10) under $G_{422}$ is given by
\be
A(45)=A(15,1,1)+A(1,3,1)+A(1,1,3)+A(6,2,2).
\ee
The contents of the gauge multiplets $A(15,1,1), A(1,3,1)$ and $A(1,1,3)$ in SM are respectively
\bea
A(15,1,1)&=&A(8,1)_0+A(3,1)_{4/3}+A(\bar{3},1)_{-4/3}+A(1,1)_0, \\ \nonumber
A(1,1,3)&=&A(1,3)_0,\\ \nonumber
A(1,3,1)&=&A(1,1)_2+A(1,1)_{-2}+A(1,1)_0.
\eea
When the neutral component $H(1,1)_0$ of the {\bf 16} Higgs fields develops a vev $<H(1,1)_0>=v^{\prime}$ $G_{422}$ 
will be broken down to SM. Then the gauge multiplets, $A(1,1)_0$ in A(15,1,1) and $A(1,1)_0$ in
 A(1,3,1), will mix with each other. That is, we need to identify the weak hypercharge $Y$ generator 
as a linear combination of the generators of $SU(4)_C\times SU(2)_R$  sharing the same quantum numbers. Using this, we can determine the 
$U(1)_Y$ term in the gaugino mass expression in function of the coupling constants $g_4$, $g_2$ corresponding 
to $SU(4)_C$ and $SU(2)_R$ respectively. Here, as we mentioned in the introduction, we assume that the 
intermediate breakings down to the SM take place all around the $M_{GUT}$ scale motivated by the MSSM 
successful unification of gauge couplings, and so we have $g_2 \sim g_4 \sim g$ leading finally to gaugino 
masses  $M_a$(a=3,2,1) in the ratio $1:-\frac{3}{2}:-1$.

\subsubsection{$SO(10)\mapsto H=SU(2)\times SO(7) \mapsto SU(3)\times SU(2)\times U(1)$} 
The first stage of this breaking chain is achieved by the $\mathbf 54$ traceless $\&$ symmetric $2^{nd}$ 
rank tensor with the non-zero vev
\begin{eqnarray}
<{\bf 54}>&=& v Diag(7/3,7/3,7/3,-1,-1,-1,-1,-1,-1,-1)
\end{eqnarray}
where the indices $1$,$2$,$3$ correspond to $SO(3)\simeq SU(2)$ and $4,\ldots 0$ correspond to SO(7). 
Subsequently SO(7) is broken to $SO(6)\simeq SU(4)$ which in turn is broken to $SU(3)\times U(1)$. As a
 result we get the gaugino masses $M_a$(a=3,2,1) in the ratio $1:-\frac{7}{3}:1$.

\subsection{The representation ${\bf 210}$}
The irrep $\mathbf 210$ of SO(10) can be represented by a $4^{th}$-rank totally antisymmetric tensor 
$\Delta_{ijkl}$. It can break SO(10) in different ways \cite{slansky81}.

\subsubsection{$SO(10) \mapsto G_{422} \mapsto SU(3)\times SU(2)\times U(1)$}
If the only non-zero vev is  $\Delta_{7890}=v$ where the indices 1 to 6 correspond to SU(4) 
while those of 7 to 0 correspond to SO(4) then the intermediate stage is $H=G_{422}$~\cite{aulakh83}. 
We see immediately here that when SU(4) would be broken to SU(3)
 we shall get massless gluinos (SU(3) gauginos). One can also see that if the only non-zero vevs are 
assumed to be $\Delta_{1234}=\Delta_{3456}=\Delta_{1256}=w$ then SO(10) is broken to $G_{3221}\equiv 
SU(3)\times SU(2) \times SU(2) \times U(1)$ \cite{aulakh83} leading, eventually, to $SU(2)_L$ massless gauginos.
 Consequently one can, in principle, assume both $v$ and $w$ $\neq 0$ and get $G_{3221}$ as an intermediate 
stage without getting massless gauginos. We would like to note here that one should not swiftly drop the 
massless gluino scenario since it is not completely excluded
phenomenologically \cite{lg} and particularly 
because, as we said earlier, the breaking could be achieved in principle from any linear combination of the 
irreps ($\mathbf 1,54,210,770$). We did not study the case corresponding to $v,w \neq 0$, neither the case 
where the intermediate stage is $G_{3211}\equiv SU(3) \times SU(2) \times U(1) \times U(1)$ achieved by 
$\Delta_{1278}=\Delta_{1290}=\Delta_{3478}=\Delta_{5678}
=\Delta_{5690}=v$ \cite{sato96} . Rather we concentrated on  
the phenomenologically interesting case of `flipped' SU(5).

\subsubsection{$SO(10) \mapsto H_{51}\equiv SU(5) \times U(1) \mapsto SU(3)\times SU(2)\times U(1)$}

The `flipped' SU(5) model \cite{barr82} exhibits some very suitable features such as fermion 
and Higgs-boson content, the natural doublet-triplet mass splitting mechanism and, among others, no cosmologically embarrassing phase transitions.

The $\bf 210$ irrep can break SO(10) to $H_{51}$ when its singlet under $H_{51}$ takes a non-zero vev which 
amounts to its non-zero components as a $4^{th}$-rank totally antisymmetric tensor  being \cite{he90}
\bea
 \Delta_{1234}&=&\Delta_{1256}=\Delta_{1278}=\Delta_{1290}=
\Delta_{3456} 
=\Delta_{3478}\\\nonumber &=&\Delta_{3490}=\Delta_{5678}=\Delta_{5690}=
\Delta_{7890}=v.
\eea 
Next we should break $H_{51}\equiv SU(5)\times U(1)_X$ to the SM$\equiv SU(3)_C\times SU(2)_L \times U(1)_Y$. 
The SU(5) group can be decomposed into $SU(3)_C\times SU(2)_L \times U(1)_Z$. The weak hypercharge $Y$ must 
be a linear combination of $Z$ and $X$. Under $SU(5)\times U(1)_X$ the {\bf 16} Higgs decomposes as
\be
{\bf 16}={\bf 10}_1+{\bf \bar{5}}_{-3}+{\bf 1}_5,
\ee
where the number on the lower right denotes the quantum number X of $U(1)_X$. Because the {\bf 10} rep. has the following 
branching rule under $SU(5) \supset SU(3)\times SU(2) \times U(1)_Z$
\[ {\bf 10} = (\bar{\bf 3},{\bf 1})_{-\frac{2}{3}}+({\bf 3},{\bf 2})_{\frac{1}{6}}+({\bf 1},{\bf 1})_1, \]
if the ${\bf 10}_1$ ( more precisely, the neutral component ({\bf 1,1}) of $SU(3)\times SU(2)$ in the ${\bf 10}_1$ rep. ) in the {\bf 16}  gets
a non-zero vev, then $H_{51}$ will break to the SM with $Y/2=\frac{1}{5} (X-Z)$ where 
$Z$ is the generator of $SU(5)$ which commutes with the generators 
of $SU(3)_C\times SU(2)_L$ and is normalized (for the five-dimensional representation of $SU(5)$ ) as
\begin{eqnarray}
Z&=&Diag(-\frac{1}{3},-\frac{1}{3},-\frac{1}{3},\frac{1}{2},\frac{1}{2}).
\label{Znormalization}
\end{eqnarray}
Note that had we chosen $\frac{Y}{2}=Z$ corresponding to the Georgi-Glashow SU(5) we would get universal 
gaugino masses.

Introducing the properly normalized $U(1)_Z$ and $U(1)_X$ generators $L_Z=\sqrt{\frac{3}{5}}Z$, 
 $L_X=\frac{1}{\sqrt{80}}X$ such that $Tr(L_Z)^2=\frac{1}{2}$ in the defining represenation of 
SU(5) and $Tr_{\bf 16}(L_X)^2=1$ in the $\bf 16$ spinor representation of SO(10)
 \cite{slansky81} we could identify the properly normalized $U(1)_Y$ field as a linear 
combination of the $U(1)_X$ and $U(1)_Z$ fields. Again, we assume that the breaking of the 
intermediate stage $H_{51}$ happens at the $M_{GUT}$ scale 
resulting in $g_1 \sim g_5 \sim g$ for the coupling constants. Then we finally get  
gaugino masses $M_a$(a=3,2,1) in the ratio $1:1:-\frac{96}{25}$.

\subsection{Summary}
We summerize in Table \ref{table1} our results for the relative gaugino masses at $M_{GUT}$ scale and 
$m_Z$ scale recalling that $M^0_3:M^0_2:M^0_1$ at $M_{GUT}$ ($M^0_a\equiv M^{GUT}_a$) evolves approximately 
to $M_3:M_2:M_1 \sim 7M^0_3:2M^0_2:M^0_1$ at the weak scale $m_Z$. The cases B, C, D correspond to different 
breaking chains respectively. The case A corresponds to the mSUGRA, i.e., the universal gaugino mass case.

\section{Phenomenological Analysis and Mass Spectra}

The gaugino mass patterns we have obtained are of phenomenological interest. 
There can, in principle, be various terms in the superpotential and Kahler potential that may give rise 
to non-universal sqaurk and slepton masses, and whether such terms exist is model-dependent. For simplicity,
we assume universal soft sfermion masses and trilinear couplings in our numerical analysis  in order to clarify the phenomenological
implications of non-universal gaugino masses. 
We use the event generator 
ISAJET \cite{baer2000} (version 7.48) to simulate models with non-universal 
gaugino mass parameters at the scale $M_{GUT}$ in this section. The model 
parameter space used in our work is expanded by $m_0$, 
$M^0_3\equiv m_{1/2}$, $A_0$, $\tan{\beta}$ and sign($\mu$). $M^0_2$ and 
$M^0_1$ can then be calculated in terms of $M^0_3$ according to Table 
\ref{table1}. ISAJET calculates an iterative solution to the 26 RGEs and 
imposes the radiative electroweak symmetry breaking constraint. This 
determines all the sparticle masses and mixings and can calculate the 
branching fractions for all sparticles, particles and Higgs bosons.

The constraints of lower bounds of sparticle and
Higgs bonson masses\cite{pdg2000} 
are included.  And we require the gauge coupling unification at the scale 
$M_{GUT}=2.0 \times 10^{16}$ GeV. Throughout the work we take $m_t=175$ GeV.

A check for the compatibility of the models with the $b \rightarrow s\gamma$ 
constraint is included. The prediction of the 
$b \rightarrow s \gamma$ decay branching ratio \cite{bertolini91} should be 
within the current experimental bounds \cite{cleo99}
\[ 2\times 10^{-4}\,<\,BR_{exp}(b\rightarrow s\gamma)\,<\,4.2 \times 10^{-4}.\]
Because there is no full next-to-leading order (NLO) formula available in SUSY models we use the 
leading order (LO) calculation with about $\pm \% 30$ theoretical uncertainty included.
This constraint is very strong for negative mu-term ($sign(\mu)=-1)$ 
\footnote{We follow the conventional definition of the sign($\mu$) 
\cite{haber85}} due to the constructively interference of SUSY contributions with the SM contributions
~\cite{cbsgamma}. It leads to a rather large (but still smaller than 1Tev) sparticle mass spectrum. For  
$sign(\mu)=+1$ , there are regions of the parameter space where the mass spectrum is low while $\tan\beta$ is 
large since the SUSY contributions destructively interfer with the Higgs's and W's contributions in this case.

The $(g-2)$ constraint of the muon anomalous magnetic dipole moment 
$a_{\mu}\equiv \frac{1}{2}(g-2)_{\mu}$ \cite{martin2001} is also considered. 
The current data of the E821 experiment \cite{brown2001} 
give the following bound on the supersymmetry contribution to $a_{\mu}$
\be 11\times 10^{-10}\,<\,  a^{SUSY}_{\mu} \,< \, 75 \times 10^{-10} \label{am} \ee
It is well-known that  the diagram with chargino-sneutrino in the loop gives a dominant contribution to 
$a^{SUSY}_{\mu}$ for general SUSY mass parameters and the muon chirality can be flipped by the Yukawa 
coupling of muon which is proportional to $1/cos\beta (\sim \tan\beta$ in the large $\tan\beta$ case)
~\cite{lnw,amuon}. Therefore, even with a relatively large mass spectrum, the bound on the 
supersymmetry contribution to $a_{\mu}$, Eq. (\ref{am}), can be satisfied in the large $\tan\beta$ case.
 On the contrary, the bound requires scenarios of small charginos and sneutrino masses when 
$\tan\beta$ is small. So the combined consideration of $(g-2)_{\mu}$ and $b\rightarrow s\gamma$
leads to that the regions of large $\tan\beta$ and low mass spectrum which are allowed by 
$b\rightarrow s\gamma$ alone decrease significantly.

Table \ref{table2} illustrates the numerical results of the mass spectra evaluated at the mass scale 
$m_Z$ for the values $m_{1/2}=300$GeV, $m_0=400$ GeV, $A_0=350$ GeV and we 
have taken a large value for the $\tan{\beta}=20$ since this would enhance 
the ($g-2$) constraint. We see from the table that the mass spectrum is relatively heavy, 
which comes in order to satisfy the $b\rightarrow s \gamma$ constraint in the $sign(\mu)= -1$ 
case, as pointed out above. All cases have neutralino LSP. Cases A, C and D have chargino NLSP, 
and case B an stau NLSP.  The four cases are experimentally distinguishable, because the 
sparticle mass splitting patterns are quite different among the four.  

\begin{figure}
\centerline{\epsffile{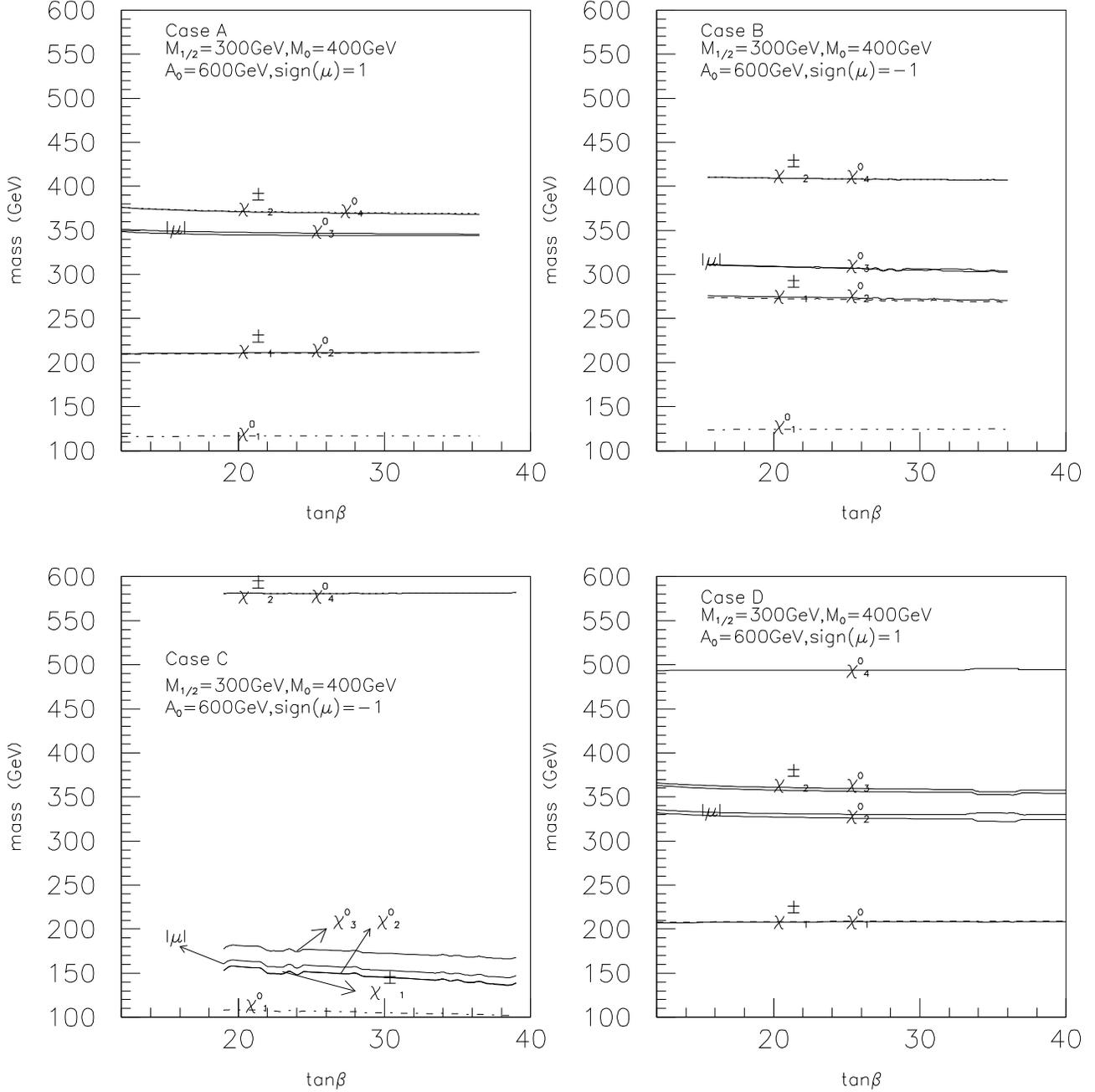}}
 \caption{The Neutralino and the Chargino masses as a function of 
  $\tan{\beta}$ for the cases in Table (\ref{table1}). Also plotted is 
  $|\mu|$.  We have taken $m_0=400$GeV, $A_0=600$ GeV and $M^0_a=m_{1/2}$ 
  times the number appearing in Table (\ref{table1}), with $m_{1/2}=300$GeV.}
 \label{figure1}
\end{figure}

In Fig. \ref{figure1} the $\tan{\beta}$ dependence of the $|\mu|$, 
neutralinos and charginos masses for the four cases in Table \ref{table1} is 
presented, where we have taken $m_{1/2}=300$ GeV, $m_0=400$ GeV, $A_0=600$ 
GeV. We noted that for far larger values of $m_0$, $m_{1/2}$ resulting in 
larger masses for the smuon and charginos, the $(g-2)_{\mu}$
constraint would be 
violated. Also we have chosen a rather large value for the trilinear scalar 
coupling $A_0$ which appears in the off-diagonal elements of the squark mass 
matrix in order to favor a large stop mass splitting. One can see from the figure that the $|\mu|$, 
neutralinos and charginos masses are insensitive to $\tan\beta$ when $\tan\beta$ is relatively
large (say, larger than 10).

\begin{figure}
\centerline{\epsffile{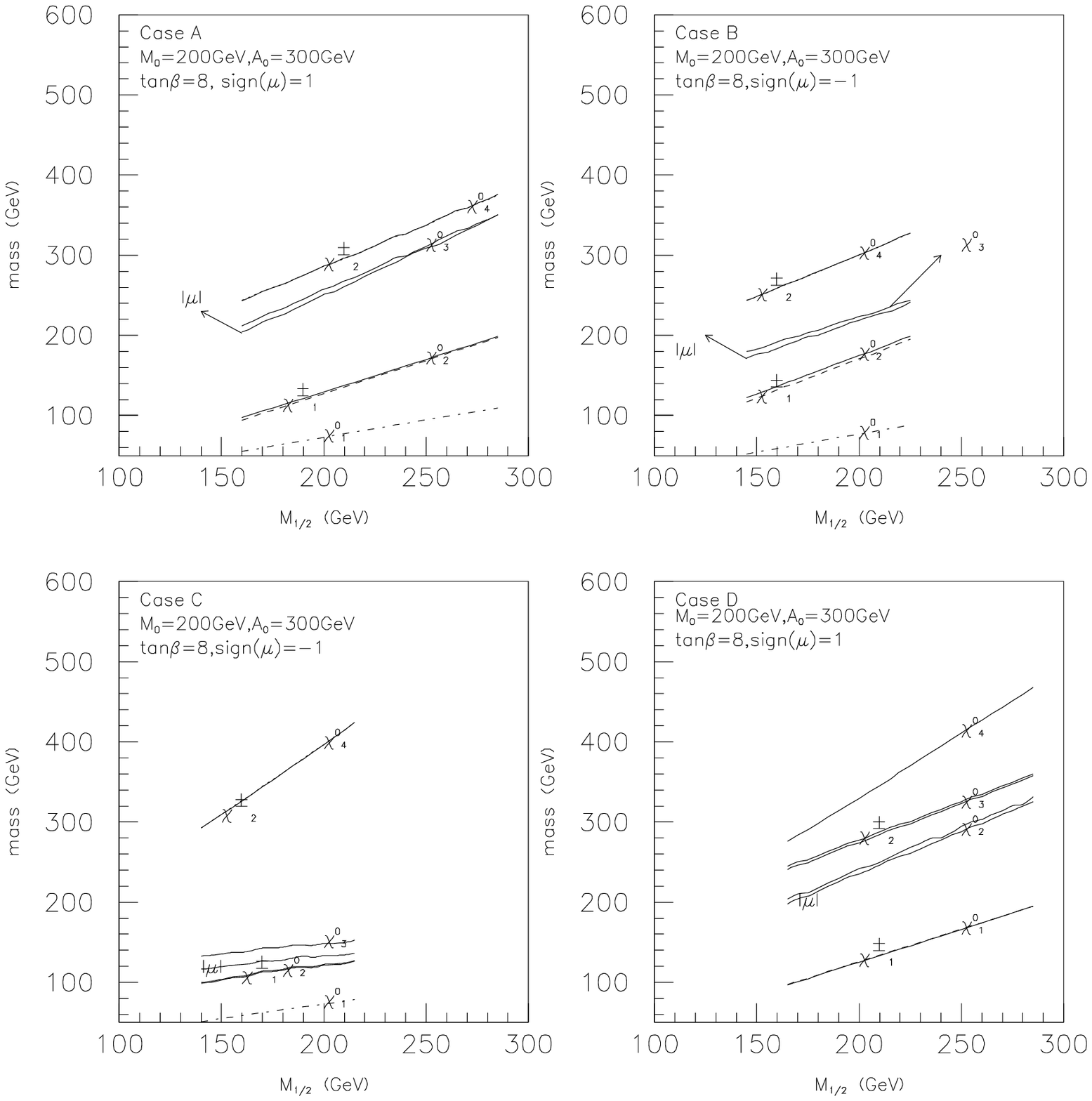}}
 \caption{The Neutralino and the Chargino masses as a function of $m_{1/2}$ 
  for the cases in Table (\ref{table1}). Also plotted is $|\mu|$.  We have 
  taken $\tan\beta=8$, $m_0=200$GeV and $A_0=300$ GeV.}
 \label{figure2}
\end{figure}

\begin{figure}
\centerline{\epsffile{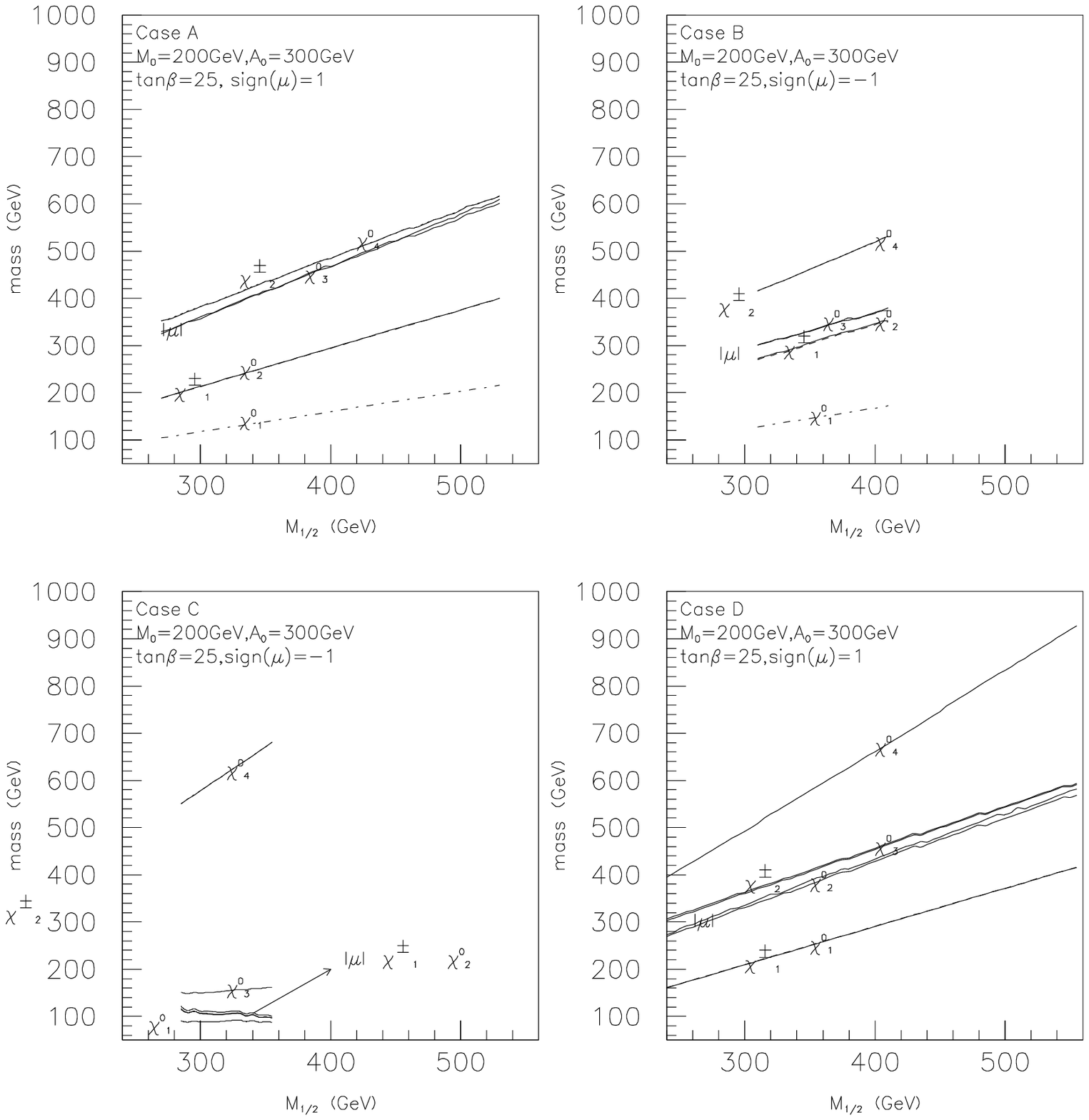}}
 \caption{The Neutralino and the Chargino masses as a function of $m_{1/2}$ 
  for the cases in Table (\ref{table1}). Also plotted is $|\mu|$.  We have 
  taken $\tan\beta=25$, $m_0=200$GeV and $A_0=300$ GeV.}
 \label{figure3}
\end{figure}

Figs. \ref{figure2} and \ref{figure3} present the $m_{1/2}$ dependence of 
the $|\mu|$, neutralinos and charginos masses for the four cases with 
$\tan\beta$ being taken to be $8$ and $25$, respectively.  
In the cases B and C corresponding to the representation $\bf 54$, the ($(g-2)_{\mu}$) 
constraint was not respected for $\mu>0$. This is in agreement with the anlysis in ref.
\cite{lnw,kmy}. As 
 pointed out in ref.\cite{lnw,kmy},  in most of the parameter space the sign of the SUSY contributions to 
$a_{\mu}$ is directly correlated with the sign of the product $M_2~\mu$ such that it is positive (negative) 
for  $M_2~\mu>0$ ($M_2~\mu<0$.  Thus, the latest result of the E821 experiment, eq.(\ref{am}), suggests 
that $M_2~\mu>0$ so that one has $\mu<0$ since $M_2<0$ in the cases B and C. In all four cases we see that 
the LSP is the neutralino but in case D corresponding to the $\bf 210$ 
representation the lightest chargino and neutralino are approximately 
degenerate while for the other three cases the approximate degeneracy 
happens for the heaviest ones.Thus, in the case D the lightest chargino is long-lived.  Therefore,
the experimental signals for the case D are different from those expected from conventional R-parity conserved
SUSY modles, e.g., mSUGRA (i.e., the case A) which have been studied 
in Ref. \cite{drees}.

\section{Conclusion}

We have studied SUSY SO(10) models in which the gaugino masses are not 
universal at the GUT scale and we have performed the group theory methods 
required to calculate their ratio. Then, for some specific values of the soft mass
parameters which are chosen to respect the experimental constraints coming from the direct search of sparticles,
$b\rightarrow s \gamma$ and $a_{\mu}$, we compared phenomenologically these models. The mass spectrum in
the case D is particularly interesting due to the presence of the approximately degenerate lightest chargino and neutralino.
 All the breaking chains allow for boundary conditions compatible with current experimental 
data on the ($b\rightarrow s\gamma$) branching ratio and the ($g-2$) 
measurement. However, these two constraints show a strong correlation and 
$a^{SUSY}_\mu$ becomes very large for the large $\tan{\beta}$ region and 
is expected to become the powerful tool in order to constrain the SUSY 
parameter space and so to decide which breaking chain is preferable.

The pattern of non-universal gaugino masses at $M_{GUT}$ is determined only by the breaking 
chain from $SO(10)$ down to SM if the scale at which the breaking of the intermediate subgroup 
happens is the same as that at the first step of the breaking chain. 
Otherwise, it also depends on the scale at which the breaking of the intermediate subgroup happens. However, the dependence is normally
weak as long as the intermediate scale is not too low \footnote{The intermediate scale is larger than about 
$10^{10}$ GeV  in most of the
model building studies (see, e.g., refs.~\cite{barr82,sato96}).}. Besides the irreps {\bf 54} and {\bf 210} of SO(10) necessary to get non-universal masses, 
we use only one more irrep of SO(10), the spinor rep. {\bf 16}, to realize the next step of breaking chains.  
This is economical in constructing a SUSY SO(10) GUT. It is important to give an explicit form of a superpotential for a SUSY SO(10)
GUT  model with non-universal gaugino masses to construct a specific model, which is beyond the 
scope of the paper and left to future work.

\section*{Acknowledgments}
This work was supported in part by the National Natural Science 
Foundation of China.  N.C. acknowledges financial support from COMSATS.

\newpage

\begin{table}
\caption{Relative masses of gauginos at the GUT scale and at the weak scale 
achieved by vevs of the $F$-term of superfields in representations 
corresponding to different breaking chains. The case A of the singlet 
representation {\bf 1} for the $F$-term corresponds to the minimal 
supergravity model.}
\label{table1}
\begin{tabular}{ccccccccc}
Case&$F_\Phi$&Intermediate Stage&$M_1^{\textrm{\scriptsize GUT}}$&
$M_2^{\textrm{\scriptsize GUT}}$&$M_3^{\textrm{\scriptsize GUT}}$&
$M_1^{m_Z}$ & $M_2^{m_Z}$ & $M_3^{m_Z}$\\
\hline
A & {\bf 1}  &                    &$1$     &$1$   &$1$ & $ 0.42$     
&$0.88$    &$3.0$\\
B & {\bf 54} & $G_{422}$          &$-1$    &$-1.5$&$1$&$-0.42$
&$-1.3$   &$3.0$\\
C & {\bf 54} & $SU(2)\times SO(7)$&$1$     &$-7/3$&$1$&$0.42$
&$-2.1$ &$3.0$\\
D & {\bf 210}& $H_{51}$           &$-96/25$&$ 1 $ &$1$&$-1.6$
&$0.88$    &$3.0$
\end{tabular}
\end{table}

\newpage

\begin{table}
\caption{Mass spectra in the four models
({\bf A}, {\bf B}, {\bf C}, {\bf D}) for $m_{1/2}=300$GeV,
$m_0=400$GeV, $A_0=350$GeV and $\tan \beta=20$.
All the masses are shown in GeV and evaluated at the scale
$m_Z$.}
\label{table2}
\begin{tabular}{rrrrrrr}
Model & $m_{{\tilde{\chi}}^\pm_{1,2}}$ & $m_{{\tilde{\chi}}^0_{1,2,3,4}}$ &
$m_{{\tilde{e}}_{1,2}}$ & $m_{{\tilde{\tau}}_{1,2}}$ & $\mu/m_{H^{\pm}}$ 
\\
 $M_{\tilde{g}}$ & $m_{H_{1,2}}$ & $m_{{\tilde{u}}_{1,2}}$ &
$m_{{\tilde{t}}_{1,2}}$ & $m_{{\tilde{d}}_{1,2}}$ &
$m_{{\tilde{b}}_{1,2}}$ \\
\hline
A & $211/375$ & $117/212/351/374$ &
$448/416$ & $394/445$&
$+348/535$ 
\\
$726$ & $114/529$&
$730/716$ & $558/704$ & $735/715$ & $657/710$ \\
\hline
B & $257/404$ & $123/259/289/404$ &
$502/416$ & $389/493$ & $-286/527$ 
\\
 $731$ & $111/521$ &
$766/717$ & $576/715$ & $770/716$ & $684/708$ \\
\hline
C & $101/579$ & $86/101/147/579$ &
$616/416$ & $385/606$& $-106/577$ 
\\
$743$ & $111/572$&
$843/718$ & $547/781$ & $847/716$ & $696/767$ \\
\hline
D & $208/358$ & $208/324/355/493$ &
$494/588$ & $482/572$& $+328/558$ 
\\
$729$ & $113/553$&
$734/765$ & $598/707$ & $738/726$ & $656/721$ 
\end{tabular}
\end{table}

\newpage

\end{document}